\documentclass[aps,prb,showpacs,twocolumn,superscriptaddress]{revtex4-2}
\usepackage{amssymb}
\usepackage{graphicx}
\usepackage{appendix}
\usepackage{dcolumn}
\usepackage{bm}
\usepackage{amsmath}
\usepackage{epstopdf}
\usepackage{xcolor}
\usepackage{float}
\usepackage{braket}
\usepackage[colorlinks,urlcolor=blue,anchorcolor=blue,linkcolor=blue,citecolor=blue,breaklinks=true]{hyperref}
\usepackage[T1]{fontenc}
\usepackage{array}
\usepackage{booktabs}

\begin{document}
%\title{Bosonic fractional Chern insulators at higher filling factors in multiple topological flat bands systems}% Force line breaks with \\
\title{Interaction-driven phase transition between bosonic Moore–Read and Laughlin states in a multiple flat-band model}

\author{Licheng Wang}
%\thanks{These authors contributed equally to the work.}
\affiliation{National Laboratory of Solid State Microstructure, Department of Physics, Nanjing University, Nanjing 210093, China}

\author{Jia-Yuan Wu}
%\thanks{These authors contributed equally to the work.}
\affiliation{National Laboratory of Solid State Microstructure, Department of Physics, Nanjing University, Nanjing 210093, China}

\author{Dong-Hao Guan}
\affiliation{National Laboratory of Solid State Microstructure, Department of Physics, Nanjing University, Nanjing 210093, China}

\author{Ai-Lei He}
\email{heailei@yzu.edu.cn}
\affiliation{College of Physics Science and Technology, Yangzhou University, Yangzhou 225002, China}

\author{Yuan Zhou}
\email{zhouyuan@nju.edu.cn}
\affiliation{National Laboratory of Solid State Microstructure, Department of Physics, Nanjing University, Nanjing 210093, China}
\affiliation{Jiangsu Key Laboratory of Quantum Information Science and Technology, Nanjing University, Nanjing 210093, China}

\date{\today}

\newcommand*\mycommand[1]{\texttt{\emph{#1}}}
\newcommand{\red}[1]{\textcolor{red}{#1}}
\newcommand{\blue}[1]{\textcolor{blue}{#1}}
\newcommand{\green}[1]{\textcolor{green}{#1}}

\begin{abstract}
Lattice models with multiple nearly flat bands provide additional degrees of freedom for realizing fractional Chern insulators and studying competition between distinct topological orders. Here we report numerical evidence for the interaction-driven phase transition between the non-Abelian and Abelian topologically ordered states at fixed filling factor in a multiple flat-band lattice model. We construct a six-site Kekulé lattice model with two low-energy nearly flat bands whose Chern numbers are $1$ and $0$, respectively. At one boson per unit cell (a site filling of $\frac{1}{6}$), the weak two-body repulsion favors a Moore-Read state predominantly occupying the lowest band, whereas stronger repulsion favors a Laughlin-like state with substantial occupations of both lower bands. We identify the two states by the low-energy spectra, particle entanglement spectra, and many-body Chern numbers, using the real-space and band-projected exact diagonalization, supplemented by infinite density-matrix renormalization group calculations. Our results provide alternative platform to study the quantum phase transition of distinct topological orders beyond the conventional multilayer systems.
\end{abstract}
\maketitle

%\tableofcontents

\emph{Introduction}--Fractional quantum Hall (FQH) effect is a typical strongly correlated state of matter with topological order, characterized by quasiparticle excitations with fractionalized charge~\cite{Tsui1982,Wen1990}. The Laughlin state is a well-known example: its quasiparticles carry fractional charge and obey Abelian fractional statistics, and the Laughlin wave function successfully explains the fractional quantum Hall plateaus in the lowest Landau level~\cite{Laughlin1983}. The Moore-Read (MR) state extends this picture to non-Abelian topological order~\cite{MooreRead1991,ReadGreen2000}. It was proposed as a candidate for the $\nu=\frac{5}{2}$ FQH state, and it has motivated extensive studies of even-denominator plateaus and competing phases~\cite{Willett1987}. Its quasiparticles are expected to be non-Abelian anyons associated with Ising-type topological order. The braiding of non-Abelian anyons does not merely produce a phase factor; rather, it acts as a nontrivial unitary transformation on the topologically degenerate ground-state space. These properties make non-Abelian FQH states candidates for realizing topologically protected quantum information~\cite{Kitaev2003Anyons,Nayak2008Anyons}.

The existence of distinct topological orders raises a natural question: how can one topological phase evolve into another as the microscopic parameters are varied? Understanding such quantum phase transitions helps study the competition among FQH states. In contrast to conventional symmetry-breaking transitions, the transitions of topological order need not be accompanied by a change of symmetry. Conventionally, studies of quantum phase transitions between topological orders have mostly focused on bilayer systems, in which the particle distribution between layers is controlled by interlayer tunneling. As a result, the quantum phase transition occurs from the Abelian two-component Halperin state to non-Abelian one-component Moore-Read (MR) state, i.e., $(220)$-Halperin to MR transition at $\nu=\frac{1}{2}+\frac{1}{2}$ in bosonic bilayer systems~\cite{Zhu2015, Liu2016BilayerDiagram, Crepel2019}. Early studies have examined the competition between $(331)$-Halperin state and MR state at $\nu=\frac{1}{2}$ in fermionic ones~\cite{HeDasXie1993,Papic2010,Peterson2010Bilayer}. DMRG calculations provided evidence for a tunneling-driven transition from the $(331)$ state to the MR state~\cite{Zhu2016FermionBilayer}. Recently, fuzzy-sphere calculations identified a gauged Majorana critical theory in a bosonic bilayer transition~\cite{Voinea2026}. In particular, experiments have suggested transitions from Jain-sequence states to Pfaffian daughter states at fillings $\frac{8}{17}$ and $\frac{7}{13}$~\cite{Singh2024JainDaughters}, and possibly between one- and two-component regimes of the $\nu=\frac{1}{2}$ state~\cite{Singh2025OneTwoComponent}. Whether similar quantum phase transition can be observed in single-layer lattice models remains unclear.

Fractional Chern insulators (FCIs) in topological nearly flat bands provide a lattice platform to realize FQH physics beyond the Landau-level paradigm~\cite{Tang2011,Sun2011,Neupert2011,BergholtzLiu2013Review,ParameswaranRoySondhi2013Review,FCI14}. Many theoretical studies have focused on isolated, partially filled Chern bands of the lattice model~\cite{Sheng2011,Regnault2011,Wang2011BosonicFCI}. Their Berry curvature and quantum metric influence the stability of FCI states~\cite{Roy2014BandGeometry,Jackson2015GeometricStability}, and ideal-flat-band constructions make the connection to Landau level physics explicit~\cite{Wang2021ExactLandauGeometry,Wang2023ModelFCI,KapitMueller2010,Qi2011,FCI7,FCI8,FCI9}. However, interactions can mix multiple bands that lie close in energy, rendering a single-band description insufficient. Recent multiband exact diagonalization (ED) studies of moir\'e systems have shown that interaction-induced band mixing can destabilize FCIs inferred from calculations restricted to a single Hartree-Fock band~\cite{Yu2025MultibandCollapse,Li2025MultibandED}. Conversely, a bosonic $\nu=\frac{1}{2}$ Laughlin-like FCI has been proposed in a triangular lattice model through occupation of the lowest two bands at integer filling~\cite{Luo2021BosonicMultibandFCI}, demonstrating that flat bands with closely spaced energies can jointly contribute to the formation of FCIs. A series of bosonic FCIs have been proposed with substantial multiple flat bands occupation, which can be stabilized with a total Chern number and nearly ideal quantum geometry~\cite{Wang2026AlternativeNuNu}. Therefore, multiple nearly flat bands provide additional degrees of freedom for controlling the FCI state. This naturally raises a further question: is it possible to drive the quantum phase transitions of distinct topological orders by tuning interactions in the multiband system?

In this paper, we study the interaction-driven phase transition between the bosonic MR and $\frac{1}{2}$-Laughlin states in a Kekul\'e-type lattice model. Among the six bands, the two lowest bands are nearly flat, and have Chern number $C_1=1$ and $C_2=0$, respectively. Restricted in the two lower band subspace, the projected effectively low-energy model contains both the two- and three-body interactions. We fix the density at one boson per unit cell and increase the onsite two-body repulsion while keeping the three-body repulsion and the single-particle Hamiltonian unchanged. At weak two-body repulsion, the numerical diagnostics support the MR state with bosons predominantly occupying the lowest band. At strong repulsion, a bosonic $\frac{1}{2}$-Laughlin-like FCI emerges with substantial occupation of both two lowest bands. We characterize these states with the low-energy spectra, many-body Chern numbers, charge pumping, as well as the particle entanglement spectra (PES).  We establish a finite-size phase diagram in the $(U_2,U_3)$ plane using two-band projected ED, by tracking the evolution of the low-energy spectrum, topological response, excitation gaps and band occupations. We also manifest the phase transition in the unprojected space. Our results provide finite-size evidence of a topological transition between non-Abelian and Abelian topological orders at fixed filling factor, establishing a lattice platform for studying the interplay between multiband occupation and competing topological orders.

\emph{Model and methods}--We construct a Kekul\'{e}-type lattice model with six orbitals per unit cell. The model  Hamiltonian is written as
\begin{eqnarray}
    H&=&H_0+H_U, \label{eq:H}\\
	H_0 &=& -t_{11}\sum_{\langle ij \rangle} e^{i\phi_1}c_{i}^{\dagger}c_{j}
	-t_{12}\sum_{\langle ij \rangle} c_{i}^{\dagger}c_{j}  \nonumber \\
	& & -t_{21}\sum_{\langle\langle ij\rangle \rangle} e^{i\phi_{21}}c_{i}^{\dagger}c_{j}
	-t_{22}\sum_{\langle\langle ij\rangle \rangle} e^{i\phi_{22}}c_{i}^{\dagger}c_{j}  \nonumber \\
	& & -t_{31}\sum_{\langle\langle\langle ij\rangle\rangle \rangle} c_{i}^{\dagger}c_{j}
	-t_{32}\sum_{\langle\langle\langle ij\rangle\rangle \rangle} c_{i}^{\dagger}c_{j} + \text{H.c.}, \\
   H_U&=&\frac{1}{2}U\sum_i n_i(n_i-1).
    \label{eq:H0}
\end{eqnarray}
Here $H_0$ and $H_U$ are the noninteracting, and interacting parts of Hamiltonian. $t_{11}$ and $t_{12}$ denote the intra- and inter-cell hopping amplitudes for the nearest-neighbor (NN), $t_{21}$ and $t_{22}$ for the second-NN, and $t_{31}$ and $t_{32}$ for the third-NN, respectively. Bond modulation and complex hopping amplitudes provide routes to engineer band topology in honeycomb-derived lattices~\cite{KariyadoHu2017,SticletPiechon2013,Lan2023StaggeredFluxFCI}. Here the staggered phases are assigned to the selected directed bond to break time-reversal symmetry while preserving sixfold rotational symmetry. The operator $c_{i}^{\dagger}$($c_{i}$) creates (annihilates) a boson at the $i$th site, and $n_i=c^{\dag}_ic_i$. For simplicity, we restrict the local occupation to $n_i=0,1,2$ in real-space ED, which implements the three-body hard-core constraint~\cite{WeinbergBukov2019QuSpin}. Increasing $U$ penalizes double occupancy at one site.

\begin{figure}[!htb]
	\begin{center}
		\includegraphics[width=0.95\linewidth]{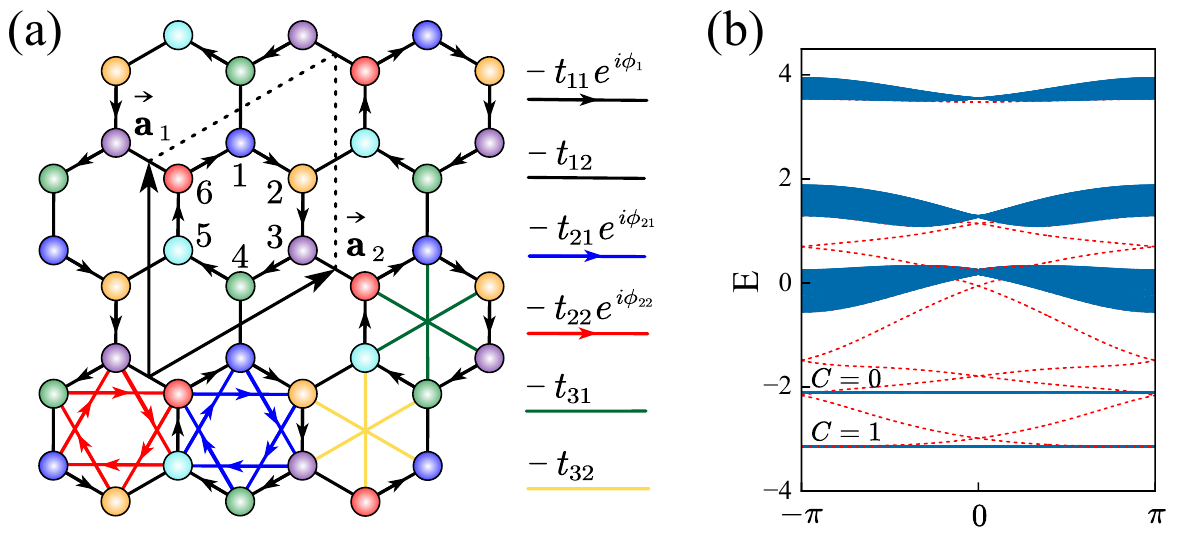}
	\end{center}
	\caption{(a) Kekul\'{e} lattice model and directed hopping terms. The $\vec{a}_1$ and $\vec{a}_2$ denote lattice vectors. Colors distinguish hopping terms, and arrows indicate staggered phases. (b) The energy spectrum with Chern number $C_1=1$ and $C_2=0$ on the cylinder geometry. Bulk and edge states are illustrated by blue solid lines and red dashed lines, respectively. }
	\label{Fig1}
\end{figure}

The original model is a six-band model. For numerical convenience, we construct the many-body basis from the lowest $M$ Bloch bands of $H_0$, with $M=1,2$.
Retaining both bands allows FCI states with appreciable multiband occupation~\cite{Luo2021BosonicMultibandFCI,Li2025MultibandED}.
The projected annihilation operator is
\begin{equation}
	\widetilde c_{\mathbf R\alpha} = \frac{1}{\sqrt{N}}\sum_{\mathbf k}\sum_{m=1}^{M}
	e^{i\mathbf k\cdot\mathbf R} u_{\alpha m}(\mathbf k)\gamma_{m\mathbf k}.
	\label{eq:projected_field}
\end{equation}
Here $N=N_x\times N_y$ is the number of unit cells, $\alpha$ labels the six orbitals within a unit cell, $u_{\alpha m}(\mathbf k)$ is the Bloch eigenvector of $H_0$, and $\gamma_{m\mathbf k}$ annihilates a boson in band $m$. The projected Hamiltonian is
\begin{equation}
	\begin{aligned}
		H_{\rm proj}={}&
		\sum_{\mathbf k}\sum_{m=1}^{M} \epsilon_m(\mathbf k) \gamma^\dagger_{m\mathbf k}\gamma_{m\mathbf k}\\
		&+\frac{U_2}{2}\sum_{\mathbf R,\alpha} (\widetilde c^\dagger_{\mathbf R\alpha})^2 (\widetilde c_{\mathbf R\alpha})^2\\
		&+\frac{U_3}{6}\sum_{\mathbf R,\alpha} (\widetilde c^\dagger_{\mathbf R\alpha})^3 (\widetilde c_{\mathbf R\alpha})^3.
	\end{aligned}
	\label{eq:Hproj}
\end{equation}
We retain the band dispersion $\epsilon_m(\mathbf k)$ and interaction matrix elements within the selected bands. Here, the additional three-body repulsion is included to suppress local triple occupancy in the retained band subspace. Its inclusion is motivated by the observation that effective three-body interactions can arise when upper bands are integrated out~\cite{Maier2012EffectiveThreeBody} and can be tuned by changing the lattice parameters in optical lattices~\cite{PaulTiesinga2015ThreeBody}. We treat $U_3$ as an independent model parameter and vary it together with $U_2$ to obtain the finite-size phase diagram.

We perform ED on a finite system of $N_x \times N_y$ unit cells(total number of sites $N_s=6 \times N_x \times N_y$), and periodic boundary conditions have been considered. We denote the boson numbers as $N_B$, and fix the filling factor of flat band $\nu=N_B/(N_x \times N_y)=1$. The momentum vector $\textbf{q}=(2 \pi k_x/N_x, 2 \pi k_y/N_y)$ is labeled as $(K_x,k_y)$. Within the two-band subspace, we vary $U_2$ and $U_3$ to calculate the finite-size phase diagram.

To study the topological order and obtain additional numerical evidence beyond the ED, we employ the iDMRG algorithm on an $L_x \times L_y$ cylinder geometry with infinite length $L_x$ and finite circumference $L_y$~\cite{White1992,McCulloch2008,CincioVidal2013,Grushin2015}. The $U(1)$ charge conservation symmetry is imposed, and the maximum truncation error is kept at the $10^{-5}$ level~\cite{McCulloch2008,CincioVidal2013,Grushin2015}. In the iDMRG calculations, the local bosonic
Hilbert space is restricted to $n_i=0,1,2$.

\emph{Multiple flat bands and quantum geometry}--For hopping parameters $(t_{11},t_{12},t_{21},t_{22},t_{31},t_{32})=(1.0,-1.154,0.705,0.252,-0.60,-0.137)$ and phase factors $(\phi_1, \phi_{21}, \phi_{22})=(0.774\pi, 1.02\pi, -0.005\pi)$, the Kekul\'{e} lattice has two low-lying nearly flat bands. The edge energy spectrum is shown in [Fig.~\ref{Fig1}(b)].
The flatness of flat band is defined as $f=\Delta / W$, where $\Delta$ and $W$ denote the band gap and bandwidth. These two flat bands have $(f_1,f_2)=(94.5,111.0)$. The Chern number of these two flat bands is $(C_1,C_2)=(1,0)$, defined as $C_n = \frac{1}{2\pi} \int_{\mathrm{BZ}} \mathrm{d}^2\mathbf{k}  \mathcal{F}_n(\mathbf{k})$, where $\mathcal{F}_n(\mathbf{k})=\nabla \times \mathcal{A}_n(\mathbf{k}) $ is the Berry curvature and $\mathcal{A}_n(\mathbf{k})$ the Berry connection.

We examine the quantum geometry of each flat band and the combined two-band subspace, since Berry curvature and quantum metric influence the stability of FCIs~\cite{Roy2014BandGeometry,Jackson2015GeometricStability}. The corresponding projectors are $P_n(\mathbf{k})=|u_n(\mathbf{k})\rangle \langle u_n(\mathbf{k})|$ and $P_{12}(\mathbf{k})=P_1(\mathbf{k})+P_2(\mathbf{k})$.
For a projector $P$, the quantum metric and Berry curvature are defined as
\begin{align}
	g_{\mu\nu}^{P}
	&=\frac{1}{2}\operatorname{Tr}
	\left[(\partial_{k_\mu}P)(\partial_{k_\nu}P)\right],\\
	\mathcal{F}^{P}
	&=i\operatorname{Tr}
	\left\{P[\partial_{k_x}P,\partial_{k_y}P]\right\}.
\end{align}
Here the trace is over the orbital degrees of freedom. For $P=P_{12}$, $\mathcal{F}^{P}$ is the trace of the non-Abelian Berry curvature, whose integral gives the total Chern number $C_{12}=C_1+C_2=1$.
To quantify the deviation from the trace condition studied in ideal flat bands ~\cite{Wang2021ExactLandauGeometry,Wang2022IdealFlatbands,Wang2023ModelFCI}, we define
\begin{equation}
	D_P(\mathbf{k})
	=g_{xx}^{P}(\mathbf{k})+g_{yy}^{P}(\mathbf{k})
	-|\mathcal{F}^{P}(\mathbf{k})|,
\end{equation}
 The $C=1$ band has more uniform Berry curvature than the $C=0$ band, and both the $C=1$ band and the combined subspace show small trace-condition deviations [details shown in Fig.S1 in Supplemental Materials (SM)]. We Note that the present two lowest flat bands with $(C_1,C_2)=(1,0)$ differ from the degenerated $C_1=C_2=1$ flat bands in conventional bilayer systems studied previously~\cite{Zhu2015, Liu2016BilayerDiagram, Crepel2019,Voinea2026}.

\begin{figure}[!htb]
	\begin{center}
		\includegraphics[width=1.0\linewidth]{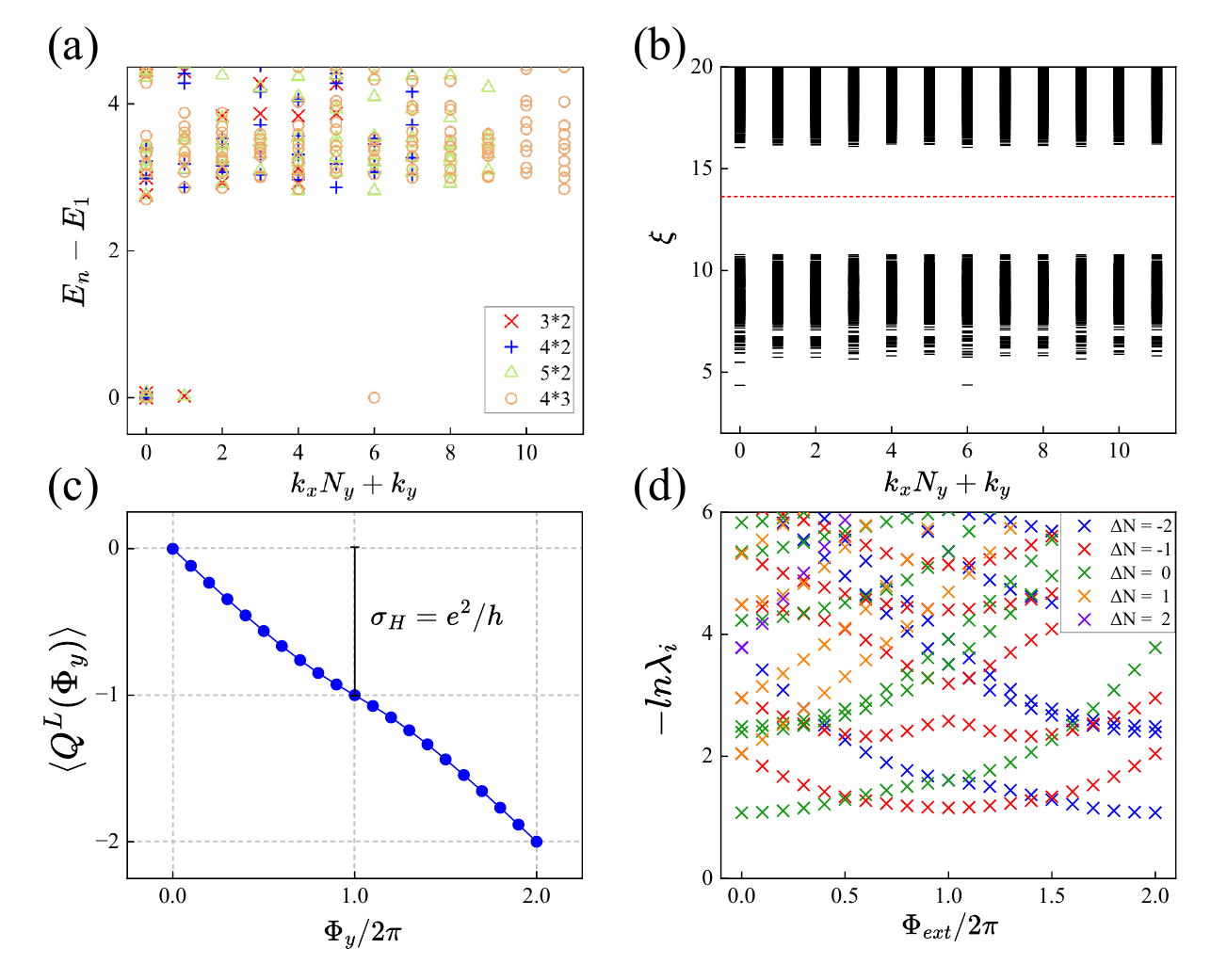}
	\end{center}
	\caption{ MR limit diagnostics. (a) and (b) Results in projected single-band subspace with parameters $M=1$, $U_2=0$, and $U_3=600$. (a) Low-energy spectra for lattice size of $N_x \times N_y$. Each spectrum are measured relative to their lowest energy. (b) PES for $4 \times3$ with $N_b=12$ and $N_A=6$. The dashed line lies in  the entanglement gap and separates the $3430$ levels.  (c) and (d) Results in unprojected Hamiltonian with $U=0$. (c) Charge pumping after two flux quanta insertion on a infinite-cylinder with $L_y=3$ and bond dimension $\chi=1200$. (d) Entanglement spectrum flow as a function of flux, five charge sectors $\Delta N$ are marked by distinct colors.}
	\label{Fig2}
\end{figure}

\emph{MR state limit}--The first flat band with $C=1$ has relatively smooth curvature and a small trace-condition deviation [Fig.S1 in SM], making it possible to realize the correlated topological states. We first consider interacting bosons at integer filling on the lowest flat band ($M=1$). At $U_2=0$ and $U_3=\infty$ limit, strong three-body repulsion produces a low-energy manifold of three states [Fig.~\ref{Fig2}(a)]. This threefold degeneracy is consistent with the bosonic MR state on a torus~\cite{Wang2012NonAbelian,FCI10}. For an isolated manifold $M$ containing $q$ orthonormal many-body states $\{|\Psi_a\rangle\}_{a=1}^{q}$, the total many-body Chern number is defined as $C_M=\frac{1}{2\pi}\int_0^{2\pi}d\theta_x\int_0^{2\pi}d\theta_y\,F_M(\theta_x,\theta_y)$, where $F_M=\mathrm{Im}\sum_{a=1}^{q}\left(\langle\partial_{\theta_y}\Psi_a|\partial_{\theta_x}\Psi_a\rangle-\langle\partial_{\theta_x}\Psi_a|\partial_{\theta_y}\Psi_a\rangle\right)$ is the trace of the non-Abelian Berry curvature over the selected manifold in boundary-twist space~\cite{NiuThoulessWu1985,Fukui2005}. These three ground states have a total Chern number $C_M=-3$, thus the average Chern number of each ground state is $C_{ave}=-1$, consistent with the Hall response expected for the bosonic MR state. Its twist-resolved Berry curvature is shown in SM. The Chern number provides the non-trivial bulk response, whereas the PES tests the quasihole structure~\cite{Sterdyniak2011,BernevigRegnault2012}. We calculate the equal-weight density matrix of the selected ground state manifold, $\rho=q^{-1}\sum_{j=1}^q|\Psi_j\rangle\langle\Psi_j|$, and trace out $N_B-N_A$ particles. The $\lambda^A_j$ are the eigenvalues of $\rho_A$, the PES levels are $\xi_j=-\ln\lambda^A_j$. Their counting below the entanglement gap probes the quasihole structure. We calculate the PES of three ground states [Fig.~\ref{Fig2}(b)]. For the size of $4 \times 3$ with $N_B=12$ and $N_A=6$, $3430$ levels lie below the entanglement gap. This counting is consistent with the MR state quasihole counting. We also calculate the PES for size of $5 \times 2$ with $N_B=10$ and $N_A=5$, and 752 levels lying below the entanglement gap [Fig. S3 in SM], providing a further finite-size check. The ground state degeneracy, Chern number and PES counting support strong evidence about a bosonic MR state at $\nu=1$ on single flat band with strong three-body interactions.

\begin{figure}[!htb]
	\begin{center}
		\includegraphics[width=1.0\linewidth]{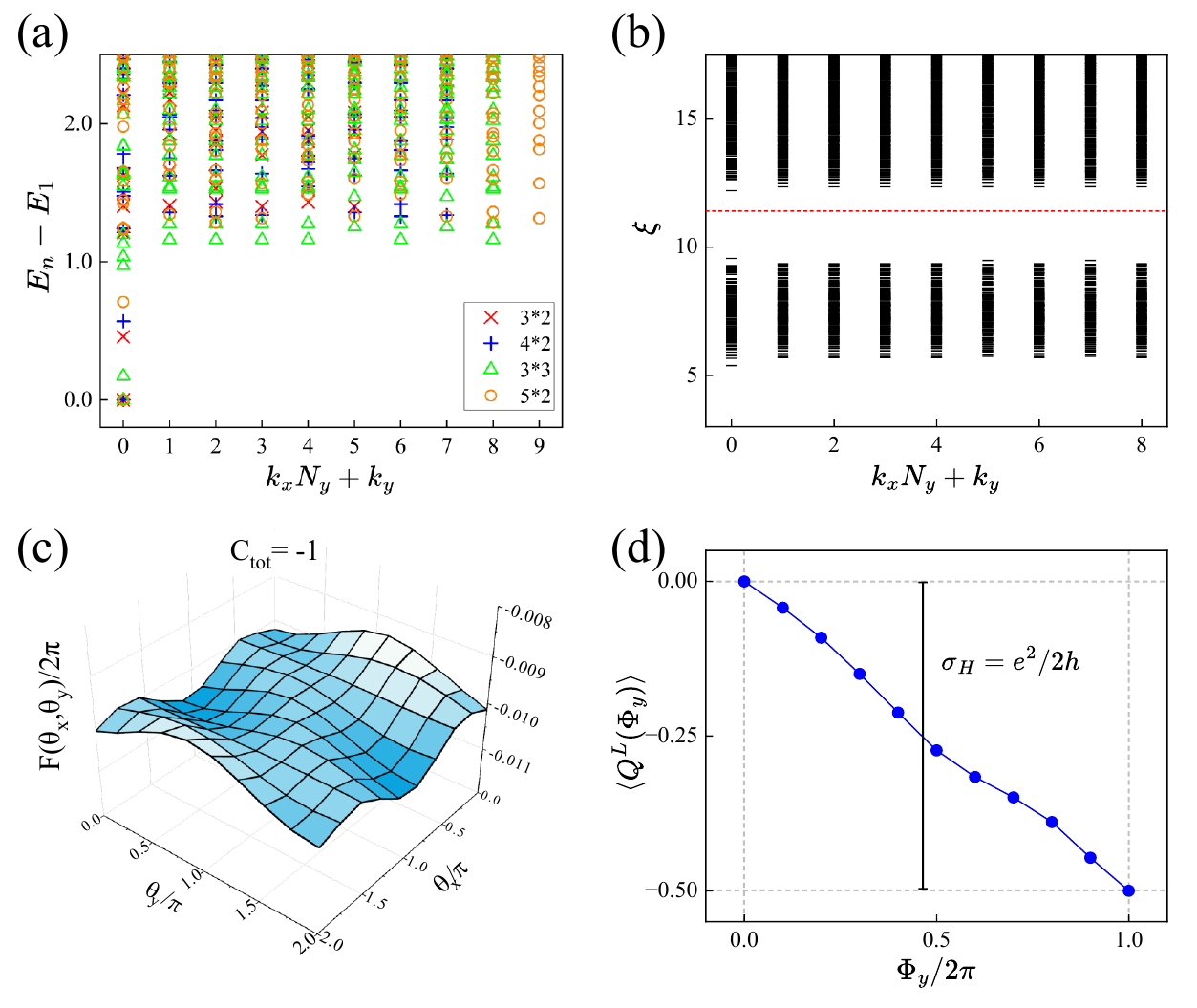}
	\end{center}
	\caption{Laughlin-like limit diagnostics. (a)-(c) Results in projected two-band subspace with parameters $M=2$, $U_2=50$, and $U_3=0$. (a) Two-band projected spectra for size of $N_x \times N_y$, measured relative to the lowest energy. (b) PES for size of $3 \times 3$, $N_B=9$ and $N_A=4$. There are $1287$ levels lie below the gap. (c) Total Berry curvature of the two ground states on a $10 \times10$ twist mesh, yielding a total Chern number $C_{12}=-1$. (d) Results in unprojected Hamiltonian with $U=12$.} Charge pumping after $2 \pi$ flux insertion with $L_y=3$ and $\chi=1200$. 
	\label{Fig3}
\end{figure}

We also examine the results on an infinite cylinder for more evidence of the topological order in unprojected Hamiltonian (Eq.~\ref{eq:H}). We obtain the charge polarization $\langle Q^L (\Phi_y) \rangle$ of the left half cylinder [Fig.~\ref{Fig2}(c)]. During the insertion of $2 \pi$ flux, the ground state wave function pumps one charge, corresponding to Hall conductivity  $\sigma_H \approx e^2 / h$. The charge pumping result agree with the many-body Chern number. As the flux is threaded, the entanglement spectrum levels evolves in the corresponding charge sectors. The entanglement flow is consistent with the identity ground state of the MR state [Fig.~\ref{Fig2}(d)]. Analogous charge pumping and entanglement flow results for the non-Abelian ground state sector at $L_y=3$ are shown in Fig.S4 in SM. We also calculate the momentum-resolved entanglement spectrum at $L_y=4$ [Fig.S5 in SM]. The low-lying sequences of degeneracy patterns show $\{1,1,3,...\}$  in even sectors and $\{1,2,4,...\}$ in odd sectors. This edge mode counting agrees with the prediction of the Identity-sector edge counting of the $SU(2)_2$ theory~\cite{LiHaldane2008,Zhu2015MRDMRG}.

To connect these results to the two-band space used below, we further calculate the MR state at weak coupling with both bands retained. For size of $4 \times 2$, $N_B=8$, $N_A=4$, $U_2=0$, and $U_3=600$, the PES contains $170$ levels below the entanglement gap [Fig. S3(b) in SM], consistent with the feature of MR state. These results support the MR state beyond the single-band projected results above.

\emph{$\frac{1}{2}$ Laughlin-like state limit}--Previous studies showed that bosonic FCIs can form in mixed subspace of nearby flat bands~\cite{Luo2021BosonicMultibandFCI,Wang2026AlternativeNuNu}. This motivates us to study whether a Laughlin-like FCI state can emerge in the present combined subspace of $C_1=1$ and $C_2=0$ bands at one boson per unit cell ($\nu=1$ for the first band). We consider both bands and introduce strong two-body repulsion with $U_3=0$.

We examine the low-energy spectra using two-band projected ED [Fig.~\ref{Fig3}(a)]. The calculation yields two quasi-degenerate low-lying ground states, consistent with the expectation for the bosonic Laughlin $\nu=\frac{1}{2}$ FCI on a torus~\cite{Wang2011BosonicFCI,Luo2021BosonicMultibandFCI}. The finite-size splitting between the two ground states depends on the lattice size, and is smallest for the size of $3 \times 3$. We next test the quasihole structure of these two ground states by PES. As shown in Fig.~\ref{Fig3}(b), the $3 \times 3$ with $N_B=9$, $N_A=4$ spectrum contains $1287$ levels below the entanglement gap. The counting is consistent with the bosonic Laughlin quasihole structure~\cite{Sterdyniak2011,BernevigRegnault2012}, providing evidence beyond the near-degeneracy of the energy spectrum. The two ground states have the total Chern number $C_{12}=-1$ [Fig.~\ref{Fig3}(c)], corresponding to a Hall response of $e^2 / 2h$, consistent with the bosonic $\frac{1}{2}$ Laughlin-like FCI state. Moreover, PES of size $4 \times 2$ with $N_B=8$, $N_A=4$ at $U_2=50$, $U_3=600$ is also obtained [Fig. S6 in SM], $660$ levels lie below the entanglement gap. This result support a Laughlin-like limit state even in the presence of the three-body repulsion, provided strong enough two-body repulsion.

 We also check the Laughlin-like state in the unprojected Hamiltonian. The iDMRG calculation shows that half a charge is pumped during $2 \pi$ flux insertion [Fig.~\ref{Fig3}(d)], consistent with this response. Fig.S7 in SM further shows the momentum-resolved entanglement spectrum of the Laughlin-like state in real-space with $L_y=5$. Its low-lying sequence $\{1,1,2,3,...\}$ matches the chiral bosonic Laughlin edge counting~\cite{Wen1995,LiHaldane2008}. The bulk Chern number, PES and edge spectrum support the Laughlin-like FCI at strong two-body interaction limit.

\begin{figure}[!htb]
	\begin{center}
		\includegraphics[width=0.9\linewidth]{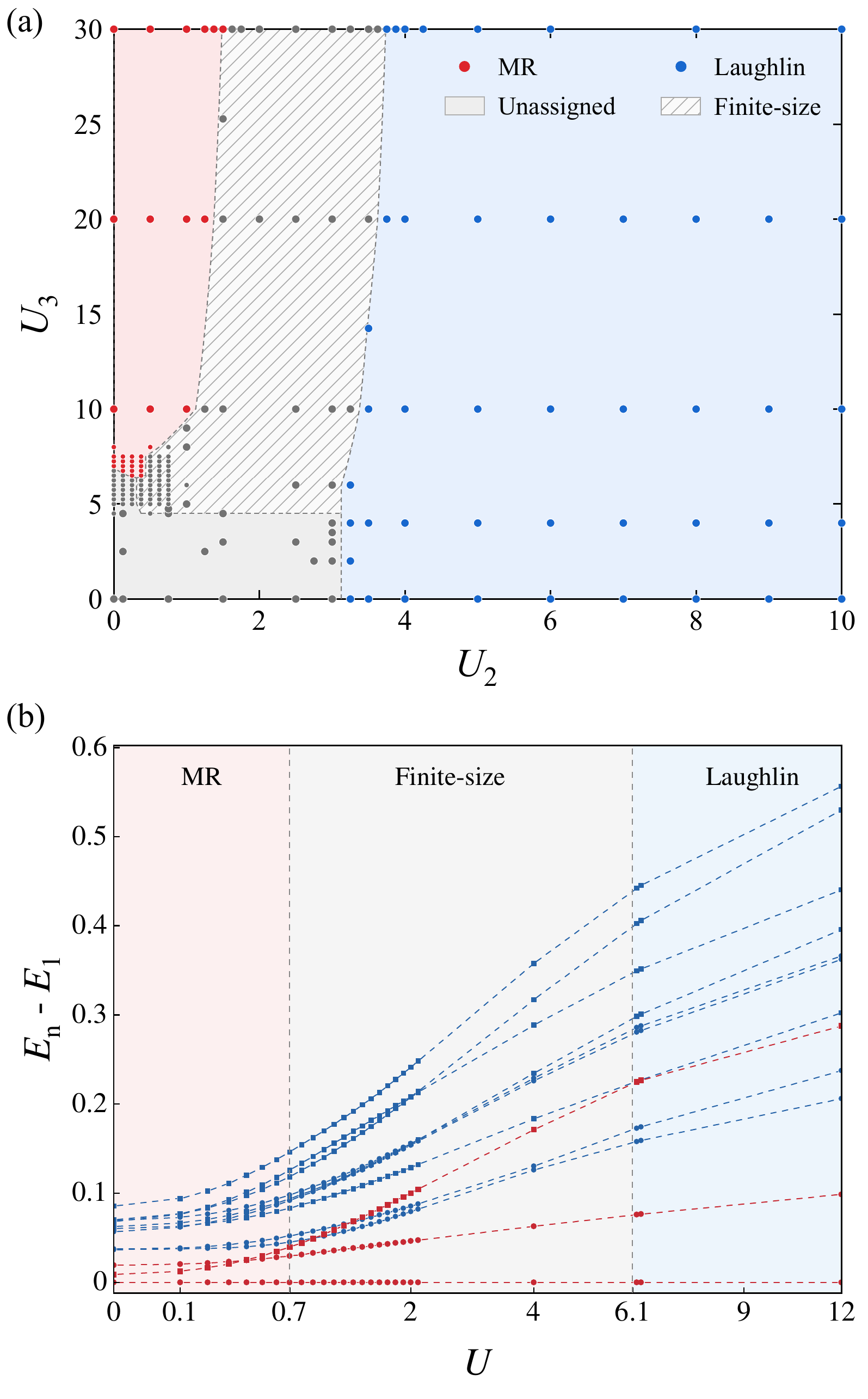}
	\end{center}
	\caption{ Interaction-driven phase diagram on a $3\times2$ lattice with $N_B=6$. (a) Finite-size phase diagram in the $(U_2,U_3)$ plane from the two-band projected ED. Red and blue indicate the MR and Laughlin regimes, respectively. Gray denotes unassigned regions, and hatching denotes a rearrangement of the low-energy spectrum. (b) Low-energy spectrum of  real-space ED as a function of the repulsion $U$ in the unprojected Hamiltonian, with local occupations restricted to $n_i=0,1,2$. Red denotes the two lowest states in the $(0,0)$ sector and the lowest state in the $(0,1)$ sector.}
	\label{Fig4}
\end{figure}

\begin{figure}[!htb]
	\begin{center}
		\includegraphics[width=1.0\linewidth]{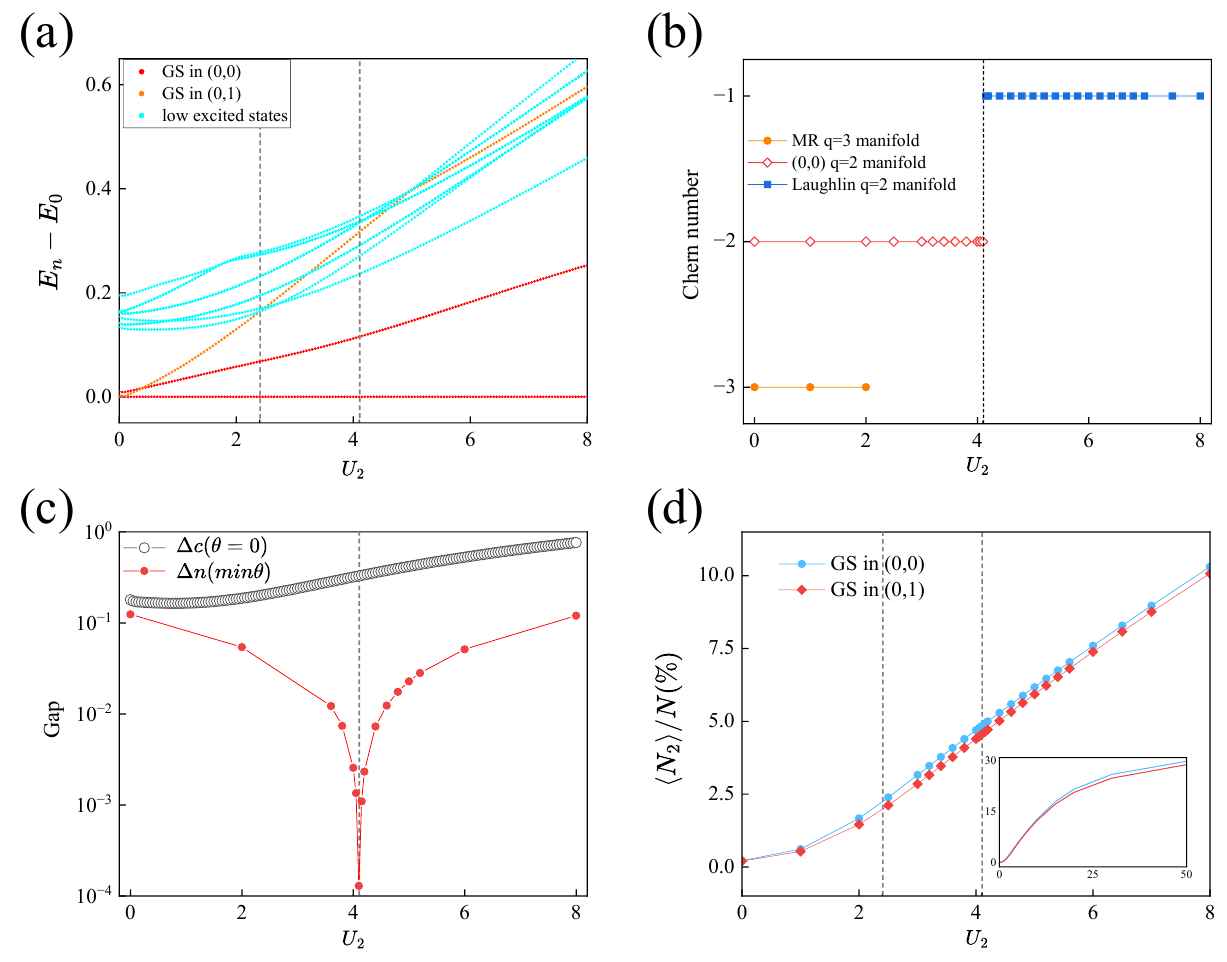}
	\end{center}
	\caption{ Two-band projected interaction scan on size of $3 \times 2$ with $N_B=6$ at fixed $U_3=600$. (a) The evolution of low-energy spectrum at twist angle $\theta=0$. Three ground states of MR state are highlighted: two lowest states in $(0,0)$ sector are illustrated in red dots, and one ground state in $(0,1)$ sector is in orange dots. (b) Chern numbers of the indicated manifolds. The orange circles show the MR state at weak $U_2$. Red diamonds and blue squares refer to the same two states in $(0,0)$ on opposite sides of the change. (c) Charge-addition gap $\Delta_c$ at $\theta=0$ and twist-resolved isolation gap $\Delta_n$ of two ground states in $(0,0)$. (d) The band occupation of the second flat band, averaged over the two states in $(0,0)$ and evaluated for the state in $(0,1)$. The inset extends the interaction range to $U_2=50$. }
	\label{Fig5}
\end{figure}

\emph{Interaction-driven evolution between topological orders}--The diagnostics of two limits support the bosonic MR state and the $\frac{1}{2}$ Laughlin state in different interaction regimes at integer filling of the first band.

 We calculate the low-energy spectra in the $(U_2,U_3)$ plane using two-band projected ED on a $3\times2$ lattice with $N_B=6$ [Fig.~\ref{Fig4}(a)]. The MR state regime appears at weak $U_2$ and strong $U_3$, while the Laughlin-like regime appears at larger $U_2$. We identify these regimes using the low-energy spectra and many-body Chern numbers, supplemented by PES with a clear counting gap. Gray shading indicates regions where the available diagnostics give no clear assignment, while the hatching denotes the intermediate regime where the low-energy spectrum undergoes finite-size rearrangements.

To follow the evolution between these states with strong three-body repulsion, we fix $U_3=600$, and vary $U_2$ within the same two-band subspace. Fig.~\ref{Fig5}(a) shows the evolution of low-energy spectrum with zero boundary twist angle. At small $U_2$, the MR manifold contains two ground state in $(0,0)$ sector and one in $(0,1)$ sector, highlighted in red and orange, respectively. As $U_2$ increases, the tracked state in $(0,1)$ rises into the excited spectrum and crosses a $(0,0)$ state near $U_2 \simeq2.4$ at $\theta=0$.

We then calculate the Chern number of the two lowest states in $(0,0)$ over the twist boundary conditions [Fig.~\ref{Fig5}(b)]. It abruptly changes from $-2$ to $-1$ near $U_2\approx 4.10$. This change does not coincide with the zero-twist level crossing. The two diagnostics probe different properties: one is the topology of selected states over all boundary twist, and the other is the level crossing ordering at a fixed boundary condition.

Twist-resolved gap closings can mediate changes in many body Chern numbers~\cite{Kourtis2017}. To track the isolation of the selected manifold, we define
\begin{equation}
	\Delta_{n}(U_2)=\min_{\bm\theta}\left[
	E_3^{(0,0)}(N_B;\bm\theta)-E_2^{(0,0)}(N_B;\bm\theta)\right].
	\label{eq:8}
\end{equation}
The minimum is obtained from the twist points. $\Delta_{n}$ is the isolation gap above the selected ground states. Near the Chern number change point, this minimum occurs at twist $(\theta_x, \theta_y)=(\pi, 0)$. The gap $\Delta_n$ closes and reopens at $U_2=4.1$ [Fig.~\ref{Fig5}(c)], manifesting the phase transition between the non-Abelian and Abelian topological orders.  For comparison, the charge addition gap, which measures the cost of adding and removing a boson on the torus, is defined as
\begin{equation}
	\Delta_c(\bm\theta)=E_0(N_B+1;\bm\theta)+E_0(N_B-1;\bm\theta)-2E_0(N_B;\bm\theta),
\end{equation}
where each $E_0$ is minimized over all momentum sectors at twist angle $\theta =0$. Throughout the scan of $U_2$, this gap remains positive and is about $0.328$ at $U_2=4.1$. The finite charge gap indicates that the system retains a finite-size signature of incompressibility near $U_2=4.1$.  In this sense, we believe that the deviation in critical value $U=2.4$ and $4.1$ mainly originates from the finite-size effect, though the intermediately incompressible phase cannot be fully ruled out. (We show the momentum-resolved density-density correlation in Fig.S8 in SM, no obvious peak is observed, which means the absence of charge-density-wave state in the intermediate regime. )

We further calculate the band occupation of the second flat band [Fig.~\ref{Fig5}(d)], which increases smoothly with $U_2$. For the two ground states in the $(0,0)$ sector, it rises from about $0.2\%$ to about $29.9\%$ at $U_2=50.0$. The increasing weight in the topologically trivial second band shows that it becomes progressively involved in forming the FCI state. At large $U_2$, this substantial occupation is consistent with the bosonic Laughlin state forming in the two band subspace. It supports a multiband interpretation of the competing states: as the two-body repulsion increases, particles redistribute between the two bands, with growing participation of the second band. The system evolves from a MR state dominated by the lowest band toward a Laughlin-like state involving substantial occupation of both bands. Such band resolved occupancy in present two band subspace resembles the layer resolved occupancy in bilayer systems studied before~\cite{Zhu2015, Liu2016BilayerDiagram, Crepel2019}, where the similar transition from the non-Abelian to Abelian topologically ordered state was also reported. However, here the single-component Laughlin state takes the place of the previous two-component Halperin state, agreeing with previously mentioned  difference of Chern number between the two-band and bilayer systems.

We also calculate the low-energy spectrum and many-body Chern number as functions of $U_2$ for the size of $4 \times 2$ [Fig.S9 in SM]. The total Chern number of the three ground states changes from $C_{123}=-3$ to $-2$ between $U_2=3.4$ and $3.5$, accompanied by a small twist-resolved gap $\Delta_{34}\simeq4.7\times10^{-5}$ near $U_2=3.4$, while the selected two ground states still have $C_{12}=-2$ on both sides of this interval. The Chern number changes to $C_{12}=-1$ between $U_2=6.3$ and $6.4$, with $\Delta_{23} \simeq5.6\times10^{-4}$ near $U_2=6.3$. The post-change Chern number is compatible with the Laughlin $\frac{1}{2}$ state. $\Delta_{j,j+1}$ denotes the minimum gap between adjacent levels on the sampled twist grid in the $(0,0)$ sector. The shift of  ground-state pair's Chern number change from $4.10-4.15$ on $3\times2$ to $6.3-6.4$ on $4\times2$ demonstrates substantial size dependence. Together with the diagnostics of two limits, these results support a finite-size change of topological order, but do not establish a thermodynamic critical coupling.

For completeness,  we show the phase diagram of the unprojected Hamiltonian [Eq.~\ref{eq:H}]. We track the low-energy spectrum on size of $3\times2$ as the repulsion $U$ increases using real-space ED [Fig.~\ref{Fig4}(b)]. The calculation uses a bosonic basis with local occupations $n_i=0,1,2$ and onsite repulsion $\frac{1}{2}Un_i(n_i-1)$ ~\cite{WeinbergBukov2019QuSpin}. At low $U\sim 0.7$, the three quasidegenerate ground states contributes a total many-body Chern number $-3$, manifesting its topological nature of MR state. With increasing $U$, one of three states of MR state ($(0,1)$ sector) moves into the low excited spectrum, leaving the intermediate regime poorly determined. Further increasing $U$, the total many-body Chern number of the remaining two ground states in $(0,0)$ abrupt changes from $-3$ to $-1$ at $U\sim 6.1$, identifying the emergence of $\frac{1}{2}$-Laughlin state. Therefore, the phase transition from the non-Abelian to Abelian topological order is also identified in the unprojected Hamiltonian.

\emph{Conclusion and discussions}--We study the interaction-driven competition between bosonic topological states in a specific multi-flat-band lattice model. Our numerical results provide strong evidence of the quantum phase transition between the Abelian and non-Abelian  topologically ordered states in both the unprojected and projected Hamiltonians. With increasing two-body interaction, the loaded bosons gradually occupy the second flat band. Accordingly, the system favors the MR state with predominantly occupying the lowest flat band at weak two-body interaction, and turns to the Laughlin state with substantially occupations of both lower bands at strong two-body interaction. We identify the phase boundary by the many-body Chern number and the particle entanglement spectra. Those results provide finite size evidence of interaction-driven band redistribution and consequent phase transition between distinct topological orders at fixed filling factor in a multiple flat band system.

The available sizes of lattice do not demonstrate whether this transition is first order or continuous, nor do they locate a thermodynamic critical point. On $3 \times 2$ lattice, the twist-resolved gap of the selected two ground states becomes very small near the Chern number change point, while the charge addition gap remains finite. The Chern number change point on $4 \times2$ lattice occurs at a different $U_2$. These results point to the reconstruction of the low-energy manifold, but do not establish the thermodynamic scaling of the neutral gap or identify the critical excitation. Earlier work studied an interlayer-tunneling-driven transition between the Abelian Halperin $(220)$ state and the non-Abelian MR state in a bosonic bilayer system~\cite{Zhu2015}. A recent work has identified a gauged Majorana critical theory for a specific bilayer transition using the fuzzy-sphere regularization~\cite{Voinea2026}. These studies provide useful references, but the present system with multiple flat bands differs in both its degrees of freedom and its tuning mechanism. Therefore, the layer-tunneling transition and critical theory cannot be inferred for the present lattice model. Establishing the order and universality class of our transition will require larger system sizes and excitation-resolved diagnostics.

A natural idea for further work is to extend this analysis to multiband subspace with different combinations of Chern numbers, for example $C_1=2$ and $C_2=0$ ~\cite{Wang2012C2FCI,Liu2012HigherC,MollerCooper2015}. Such systems would allow us to study whether interaction driven redistribution between bands can stabilize other Abelian and non-Abelian topological orders ~\cite{Sterdyniak2013}. Moreover, the effects of band geometry and interactions on these topological phase transitions can be further investigated. Given the fractional quantum anomalous Hall states reported in twisted $\text{MoTe}_2$ systems~\cite{Cai2023MoTe2,Zeng2023MoTe2,Park2023MoTe2,Xu2023MoTe2}and rhombohedral graphene superlattices~\cite{Ji-Nature2024,Lu-Nature2025-1,Lu-Nature2025-2,Xie-NM2025}, as well as the theoretical proposals for bosonic exciton fractional Chern insulators in moir\'e heterostructures~\cite{PereaCausin2025BosonicExcitonFCI}, this question is of current interest. These material systems are not direct realizations of lattice model filled with bosons, but they motivate further study of the multiband mechanisms underlying fractional topological orders.\\

\emph{Acknowledgement}--This work was supported by the National Natural Science Foundation of China (Grants No. 12204404 and No. 12374137) and the Natural Science Foundation of Jiangsu Province (Grant No. BK20231397).

\emph{Data availability}--The data that support the findings of this article are not publicly available. The data are available from the authors upon reasonable request.

\bibliography{references}
\end{document}